\documentclass[aps,prl,twocolumn,groupedaddress,showpacs]{revtex4}
\usepackage{bm}
\usepackage{graphicx}
\begin{document}

\title[Submitted to New Journal of Physics] {Electronic structure of heavily electron-doped BaFe$_{1.7}$Co$_{0.3}$As$_2$ studied by angle-resolved photoemission}

\author{
	Y. Sekiba$^1$, 
	T. Sato$^{1,2}$, 
	K. Nakayama$^1$, 
	K. Terashima$^3$, 
	P. Richard$^4$, 
	J. H. Bowen$^5$, 
	H. Ding$^5$, 
	Y.-M. Xu$^6$, 
	L. J. Li$^7$, 
	G. H. Cao$^7$, 
	Z.-A. Xu$^7$, 
	and T. Takahashi$^{1,4}$
	}
	
\address{}
\address{$^1$Department of Physics, Tohoku University, Sendai 980-8578, Japan}
\address{$^2$TRIP, Japan Science and Technology Agency (JST), Kawaguchi 332-0012, Japan}
\address{$^3$UVSOR Facility, Institute for Molecular Science, Okazaki 444-8585, Japan}
\address{$^4$WPI Research Center, Advanced Institute for Materials Research, Tohoku University, Sendai 980-8577, Japan}
\address{$^5$Beijing National Laboratory for Condensed Matter Physics, and Institute of Physics, Chinese Academy of Sciences, Beijing 100190, China}
\address{$^6$Department of Physics, Boston College, Chestnut Hill, MA 02467, USA}
\address{$^7$Department of Physics, Zhejiang University, Hangzhou 310027, China}

\date{\today}

\begin{abstract}
We have performed high-resolution angle-resolved photoemission spectroscopy on heavily electron-doped non-superconducting (SC) BaFe$_{1.7}$Co$_{0.3}$As$_2$.  We find that the two hole Fermi surface pockets at the Brillouin zone center observed in the hole-doped superconducting Ba$_{0.6}$K$_{0.4}$Fe$_2$As$_2$ are absent or very small in this compound, while the two electron pockets at the zone corner significantly expand due to electron doping by the Co substitution.  Comparison of the Fermi surface between non-SC and SC samples indicates that the coexistence of hole and electron pockets connected {\it via} the antiferromagnetic wave vector is essential in realizing the mechanism of superconductivity in the iron-based superconductors.
\end{abstract}
\pacs{74.25.Jb, 74.70.-b, 79.60.-i}

\maketitle

  The discovery of superconductivity in FeAs-based superconductors \cite{Kamihara,Takahashi} has attracted considerable interests since their transition temperatures ($T_{\rm c}$'s; maximally $\sim$55 K \cite{RenCPL,AIST}) are the highest amongst the known superconductors except cuprates.  Parent compounds of iron-based superconductors commonly show a collinear antiferomagnetic (AF) spin density wave \cite{DaiNature} with distinct anomalies in the transport and thermodynamic properties.  Doping holes or electrons into the parent compounds gives rise to superconductivity with $T_{\rm c}$ values typically above 20 K \cite{Rotter,Sefat}.  For instance, in $A$Fe$_2$As$_2$ ($A$: Alkali metals and Alkali-earth metals), so-called $122$ system, holes are doped by the chemical substitution of $A$$^{2+}$ ions by potassium ions (K$^+$), while electrons are doped by the replacement of divalent iron atoms with trivalent cobalt (Co) or tetravalent nickel (Ni) irons \cite{Sefat,Hiiramatsu}.  On the other hand, in $R$FeAsO compounds ($R$: Rare earth atoms; the $1111$ system), doping of hole carriers appears to be difficult \cite{Kamihara}.  Therefore the $122$ system provides a precious opportunity to explore the role of doping with both types of carriers with the same crystal structure.  To elucidate the electronic states relevant to the occurrence of superconductivity in the $122$ series, angle-resolved photoemission spectroscopy (ARPES) has been performed on the hole-doped Ba$_{1-x}$K$_x$Fe$_2$As$_2$ \cite{HongEPL,ZhouBK,Borisenko,Hasan} and its parent compound Ba(or Sr)Fe$_2$As$_2$ \cite{FengSr,AdamBK}.  ARPES studies of the SC samples revealed the presence of multiple Fermi surfaces (FSs) derived from the Fe 3$d$ orbitals, as well as FS-sheet-dependent SC gaps and many-body interactions \cite{HongEPL,Pierrekink}.  On the other hand, few such experiments have been carried out in the electron-doped counterpart, although this point is crucial despite its importance in examining the possibility of electron-hole symmetry or asymmetry.  As demonstrated by electrical resistivity measurements, the $T_{\rm c}$ value of electron-doped BaFe$_{2-x}$Co$_x$As$_2$ shows a maximum upon doping at around $x$ = 0.15-0.2 ($T_{c}^{mid}$ = 25.5 K), and finally disappears at $\sim$0.3 \cite{Cosample}.  Clarifying the microscopic origin of the characteristic SC phase diagram in the electron-doped iron-based superconductor would be essential in fully understanding the SC mechanism of the iron-based superconductors.  It is thus of particular importance to investigate the band structure and the FS of the electron-doped 122 compounds by performing ARPES measurements on these materials and also to directly elucidate the doping evolution of their electronic states from the SC to the metallic region.
  
	In this paper, we report high-resolution ARPES results on metallic BaFe$_{1.7}$Co$_{0.3}$As$_2$ ($T_{\rm c}$ = 0 K).  We have determined the band structure near $E_{\rm F}$ and the FS topology, and compared with the results obtained on superconducting BaFe$_{1.85}$Co$_{0.15}$As$_2$ ($T_{\rm c}$ = 25.5 K).  We demonstrate that, unlike BaFe$_{1.85}$Co$_{0.15}$As$_2$, the interband scattering condition {\it via} the AF wave vector is not satisfied in BaFe$_{1.7}$Co$_{0.3}$As$_2$.  We discuss the implications of our results in comparison with the hole-doped system.

The high-quality single crystals of BaFe$_{2-x}$Co$_x$As$_2$ used in this study were grown by the self flux method, the same growth method as for BaFe$_{2-x}$Ni$_x$As$_2$ \cite{sample}.  Co content was determined by the energy-dispersive X-ray spectroscopy.  The starting materials (nominal compositions) for BaFe$_{1.7}$Co$_{0.3}$As$_2$ and BaFe$_{1.85}$Co$_{0.15}$As$_2$ are BaFe$_{1.6}$Co$_{0.4}$As$_2$ and BaFe$_{1.8}$Co$_{0.2}$As$_2$, respectively.  The electrical resistivity of BaFe$_{1.7}$Co$_{0.3}$As$_2$ does not show any sign of superconductivity down to 2 K.  High-resolution ARPES measurements were performed using a VG-SCIENTA SES2002 spectrometer with a high-flux discharge lamp and a toroidal grating monochromator.  We used the He I$\alpha$ resonance line ($h$$\nu$ = 21.218 eV) to excite photoelectrons.  The energy and angular (momentum) resolutions were set at 4-10 meV and 0.2$^\circ$ (0.007\AA$^{-1}$), respectively.  Clean surfaces for ARPES measurements were obtained by {\it in-situ} cleaving of crystals in a working vacuum better than 5$\times$10$^{-11}$ Torr.  The Fermi level ($E_{\rm F}$) of the samples was referenced to that of a gold film evaporated onto the sample substrate.  Mirror-like sample surfaces were found to be stable without obvious degradation for the measurement period of 3 days.
	


	Figures 1(a) and (b) show energy distribution curves (EDCs) of BaFe$_{1.7}$Co$_{0.3}$As$_2$, called here the Co0.3 sample, in a relatively wide energy region with respect to $E_{\rm F}$ measured at 15 K with the He I$\alpha$ line ($h$$\nu$ = 21.218 eV) along two high-symmetry lines (a) $\Gamma$X and (b) $\Gamma$M.  In the $\Gamma$X cut (Fig. 1(a)), we find a band showing a holelike dispersion centered at the $\Gamma$ point which approaches $E_{\rm F}$ around the $\Gamma$ point.  In contrast to the hole-doped Ba$_{0.6}$K$_{0.4}$Fe$_2$As$_2$, we do not find apparent crossing of a holelike band in this energy range.  In the $\Gamma$M cut (Fig. 1(b)), we identify an electronlike band crossing $E_{\rm F}$ midway between the $\Gamma$ and M points.  To see more clearly the dispersive bands, we have mapped out the ARPES intensity as a function of wave vector and binding energy and we show the results in Figs. 1(c) and (d) for the $\Gamma$X and $\Gamma$M directions, respectively.  We also plot the results of first-principle band-structure calculations at $k_{\rm z}$ = 0 and $\pi$ (blue and red curves, respectively) \cite{ZFang}.  The calculated bands for BaFe$_2$As$_2$ are shifted downward by 90 meV, and then divided by a renormalization factor of 2.  As seen in Figs. 1(c) and (d), although some portions of the experimentally determined band structure show a rough agreement with the renormalized band calculations, most of bands show noticeable discrepancies.  For example, the higher-energy bands observed at the $\Gamma$ point around 0.2 eV in the experiment are found between 0.4 and 0.6 eV in the calculations, and the bottom of the electronlike band at the M point in the experiment is measured at $\sim$0.1 eV whereas calculations predict $\sim$0.3 eV.  Moreover, fine structures in the calculations, such as the complicated band dispersion around the M point and the appearance of a small hole pocket near the X point, are not well reproduced in the experiment.

\begin{figure}[htb]
\begin{center}
\includegraphics[width=22 pc]{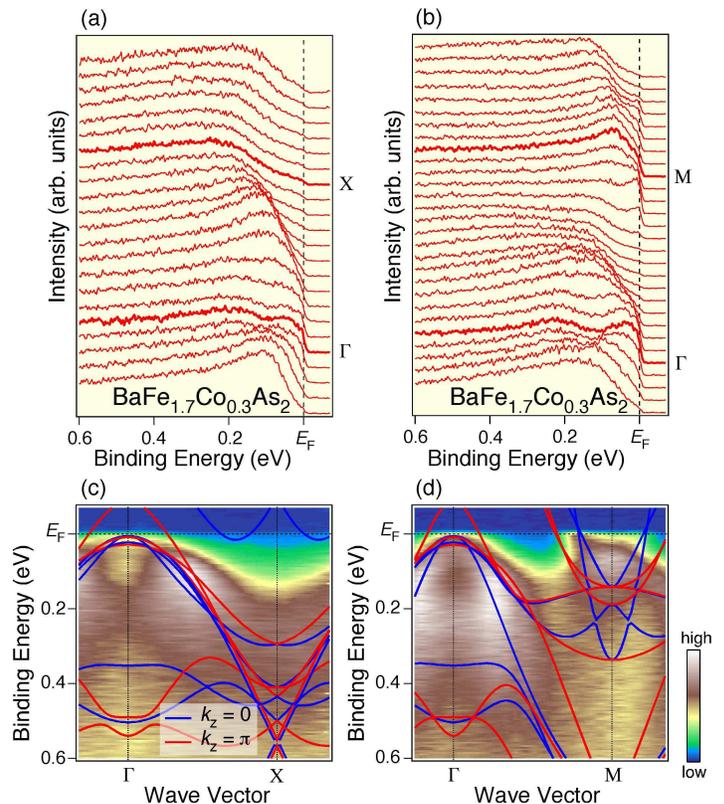}
\vspace{1mm}
\caption{EDCs near $E_{\rm F}$ of non-SC BaFe$_{1.7}$Co$_{0.3}$As$_2$ measured at 15 K with the He I$\alpha$ line ($h$$\nu$ = 21.218 eV) along two high symmetry lines (a) $\Gamma$X and (b) $\Gamma$M.  (c) and (d), ARPES intensity plots as a function of wave vector and binding energy along the $\Gamma$X and $\Gamma$M lines, together with the band dispersion from the band calculations for $k_{\rm z}$ = 0 and $\pi$ (blue and red curves, respectively).  Calculated bands for BaFe$_2$As$_2$ \cite{ZFang} were shifted downward by 90 meV and then renormalized by a factor of 2.}
\label{fig:toosmall}
\end{center}
\end{figure}
	 
\begin{figure*}[htb]
\begin{center}
\includegraphics[width=36 pc]{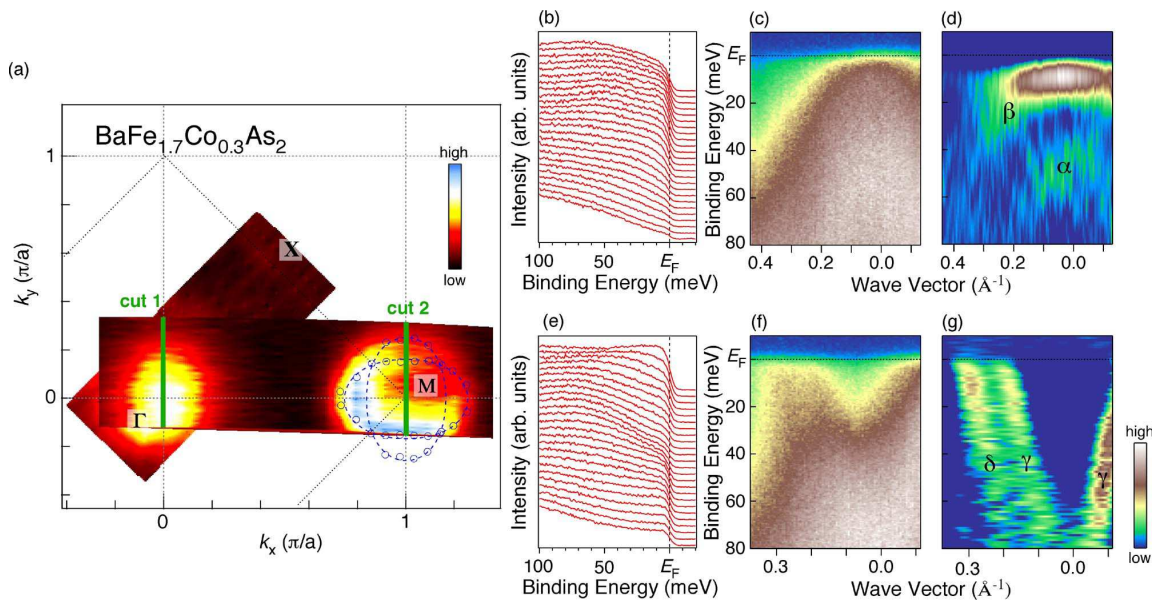}
\vspace{2mm}
\caption{(a) ARPES intensity plot at $E_{\rm F}$ of BaFe$_{1.7}$Co$_{0.3}$As$_2$ as a function of the two-dimensional wave vector measured at 15 K.  The intensity at $E_{\rm F}$ is obtained by integrating the spectra within $\pm$5 meV with respect to $E_{\rm F}$.  The location of the $k_{\rm F}$ points (open circles) has been determined by tracing the peak position of the second derivative of the MDCs at $E_{\rm F}$, and then the $k_{\rm F}$ points were folded by assuming the mirror symmetry with respect to the $\Gamma$M line ($k_{\rm x}$ axis).  The dashed lines were extracted by fitting the location of the $k_{\rm F}$ points by using the tight-binding formula \cite{Hongband, TB}.  (b) Representative EDCs in the vicinity of $E_{\rm F}$ and (c) its intensity plot as a function of binding energy and wave vector, measured along cut 1 indicated by a green line in (a).  (d) Intensity of the negative part of second derivative of EDCs along cut 1.  It is noted that the second-derivative intensity of the $\beta$ band near $E_{\rm F}$ is affected by the Fermi-edge cut-off.  (e) and (f), same as (b) and (c) but measured along cut 2.  (g) Intensity of the negative part of second derivative of MDCs along cut 2.}
\label{fig:wide}
\end{center}
\end{figure*}

	Figure 2(a) displays the ARPES intensity plot at $E_{\rm F}$ of the Co0.3 sample as a function of the in-plane wave vector measured at 15 K.  We identify two bright spots centered at the $\Gamma$ and M point.  The intensity centered at the M point originates in the electronlike band as seen in Fig. 1(b), while that at the $\Gamma$ point is produced by the spectral weight at/near the top of the holelike bands.  To examine the character of the FSs in more detail, we plot in Fig. 2(b) the EDCs in the vicinity of $E_{\rm F}$ as well as its intensity plot (Fig. 2(c)) along cut 1 ($\Gamma$M direction), measured with a higher energy resolution.  As shown in Fig. 2(d) where the second-derivative of the EDCs is plotted as a function of binding energy and wave vector, two holelike bands can be recognized.  One has a top of dispersion around 50 meV at the $\Gamma$ point, while the other is located at lower binding energy with a top of dispersion at the $\Gamma$ point at less than 10 meV.  According to a previous ARPES study of Ba$_{0.6}$K$_{0.4}$Fe$_2$As$_2$ \cite{HongEPL}, the inner and outer holelike bands are attributed to the $\alpha$ and $\beta$ bands, respectively.  Judging from the presence of a Fermi-edge cutoff around the $\Gamma$ point, the $\beta$ band may touch $E_{\rm F}$, although it is hard to distinguish the crossing point with the present experimental accuracy.  In any case, the FS produced by the $\beta$ band would be negligibly small as compared to the electron pocket at the M point, in sharp contrast to the largest size of the $\beta$ FS in the hole-doped Ba$_{0.6}$K$_{0.4}$Fe$_2$As$_2$ \cite{HongEPL,Hongband}.  Besides one electronlike FS/dispersion which is prominent around the M point as seen in Figs. 2(a) and (e), we find clear evidence for another electron pocket.  As illustrated in the second-derivative plot of the momentum distribution curves (MDCs) measured along cut 2 (Fig. 2(g)), two nearly parallel electronlike bands crossing $E_{\rm F}$ are clearly distinguished.  By comparing with the hole-doped Ba$_{0.6}$K$_{0.4}$Fe$_2$As$_2$ \cite{HongEPL,ZFang,Hongband,K1Sato,Singh,JDai}, these inner and outer FSs are attributed to the $\gamma$ and $\delta$ FSs, respectively.  As in the case of the hole-doped system \cite{Hongband}, the existence of these two electron pockets in the Co0.3 sample is basically explained by the presence of two ellipsoidal FSs elongated along two $\Gamma$M directions (along $k_{\rm x}$ and $k_{\rm y}$ axes) as indicated by dashed lines which represent the tight-binding \cite{Hongband, TB} fits of the determined $k_{\rm F}$ (Fermi vector) points (open circles) in Fig. 2(a).  It is noted here that, to determine the energy position of bands from the second-derivative plot, it is better to differentiate spectral intensity perpendicularly to the band dispersion. Hence, we have selected the EDCs method when the top or bottom of bands is included in the energy range of interest, as in the case for Figs. 2(b)-(d).  On the other hand, the MDCs method is more reliable for Figs. 2(e)-(g) since the bottom of bands is away from $E_{\rm F}$ ($\sim$80 meV) and the band dispersion is steep around $E_{\rm F}$.

\begin{figure}[htb]
\begin{center}
\includegraphics[width=16 pc]{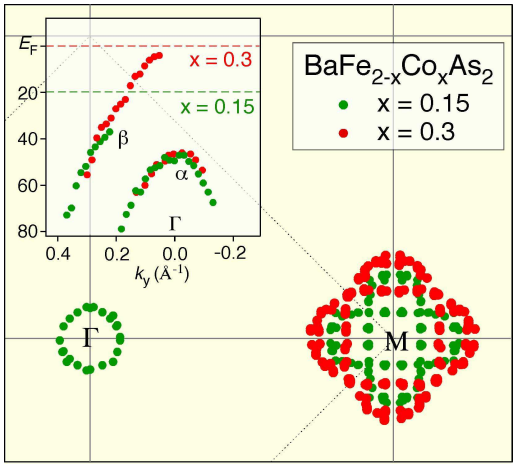}
\vspace{2mm}
\caption{(a) Comparison of experimentally determined $k_{\rm F}$ points between non-SC BaFe$_{1.7}$Co$_{0.3}$As$_2$ (Co0.3) and SC BaFe$_{1.85}$Co$_{0.15}$As$_2$ (Co0.15).  The $k_{\rm F}$ points are symmetrized by assuming the four-fold symmetry with respect to the $\Gamma$ and M points.  Inset shows the experimental band dispersion in the vicinity of $E_{\rm F}$ around $\Gamma$ point determined by tracing the peak position of background-subtracted EDCs divided by the Fermi-Dirac distribution function.  An ARPES spectrum away from the $k_{\rm F}$ points was chosen as a background.  The chemical potential of the Co0.3 sample is shifted upward by 20 meV with respect to that of the Co0.15 sample.  The peak position of the $\beta$ band near $E_{\rm F}$ for the Co0.3 sample does not coincide exactly with the second-derivative intensity in Fig. 2(d), which is affected by a Fermi-edge cut-off.}
\label{fig:toosmall}
\end{center}
\end{figure}

	Figure 3 displays a direct comparison of the momentum location of $k_{\rm F}$ points between the metallic non-SC Co0.3 and the SC Co0.15 samples.  Upon electron doping, the holelike $\beta$ FS, as observed in the Co0.15 sample \cite{Terashima0.2}, disappears or becomes indistinguishably small in the Co0.3 sample.  Simultaneously, the size of the two electron pockets at the M point significantly expands with electron doping.  Indeed, the estimated volume of the $\beta$ FS with respect to the first unfolded Brillouin zone for the Co0.3 sample is 0-0.5 $\%$, much smaller than that for Co0.15 (1.5$\pm$0.5 $\%$), and the volume of the electronlike FS is larger in the Co0.3 sample (6.5$\pm$0.5 $\%$) as compared to the Co0.15 sample (3.5$\pm$0.5 $\%$).  By taking into account the volume of all FS sheets including two electron pockets, the total electron concentration of the Co0.3 sample is estimated to be 0.13$\pm$0.01 electrons/Fe, where the error in the FS volume and carrier concentration originates in the experimental uncertainties in the determination of the location of the $k_{\rm F}$ points.  This value is close to the expected value (0.15 electrons/Fe).  We emphasize here that the FSs at the $\Gamma$ point of the SC and non-SC samples are drastically different  since the $\beta$ hole pocket observed in the Co0.15 sample is absent or fairly small in the Co0.3 sample.  This marked difference is reasonably explained by a chemical potential shift due to electron doping.  As shown in the inset of Fig. 3, the energy position of bands in the vicinity of $E_{\rm F}$ near the $\Gamma$ point for the Co0.3 and Co0.15 samples quantitatively matches when we shift down the bands of Co0.15 by 20 meV, suggesting the basic applicability of a rigid-band model.  
A simple extrapolation using a tight-binding formula suggests that the holelike $\beta$ band in the Co0.15 sample tops at less than 20 meV above $E_{\rm F}$ \cite{Terashima0.2}.  The $\beta$ band for the Co0.3 sample would be nearly completely occupied if we assume the observed chemical potential shift of 20 meV.

\begin{figure}[htb]
\begin{center}
\includegraphics[width=20 pc]{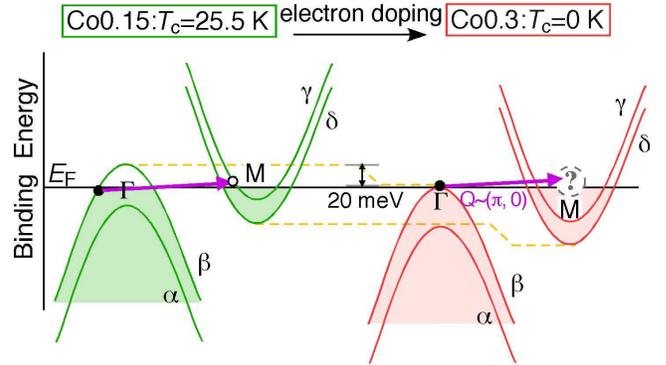}
\vspace{2mm}
\caption{Comparison of energy bands between the Co0.15 and Co0.3 samples.  The interband scattering is dramatically suppressed in the non-SC Co0.3 sample since the holelike $\alpha$ and $\beta$ bands at the $\Gamma$ point are basically occupied.}
\label{fig:toosmall}
\end{center}
\end{figure}

	Now we discuss the mechanism of $T_{\rm c}$ reduction and resultant disappearance of superconductivity in the overdoped region.  As seen in the schematic band structure in Fig. 4, the $\beta$ FS and $\gamma$ FS of the SC Co0.15 sample are well connected by the AF wave vector $Q_{\rm AF}$ = ($\pi$, 0) \cite{Terashima0.2}.  Therefore, the enhanced interband scattering {\it via} the $Q_{\rm AF}$ vector would promote the pair scattering between these two FSs by low-energy fluctuations, leading to an increase of the pairing amplitude \cite{Mazin,Kuroki,FWang}.  On the other hand, in the non-SC Co0.3 sample, this interband scattering would decrease since the $\beta$ band is in the occupied side and the size of the $\gamma$ and $\delta$ pockets expands.  This suggests that the interband scattering between the hole and electron pockets is essential to achieve high-$T_{\rm c}$ values.  The absence (deterioration) of this scattering condition in the Co0.3 sample, likely assisted by the Co-substitution-induced disorder, would completely kill superconductivity.  It is also remarked here that similar mechanism is also at work in the hole-doped Ba$_{1-x}$K$_x$Fe$_2$As$_2$ where the disappearance of the electronlike $\gamma$ band and resultant ill-defined interband scattering condition plays a critical role to the $T_{\rm c}$-reduction mechanism in the overdoped region \cite{K1Sato}.  All these experimental results suggest a possible electron-hole symmetry of the interband scattering and the pairing mechanism in the iron-based superconductors.

In summary, we have reported ARPES results on BaFe$_{1.7}$Co$_{0.3}$As$_2$ and determined the band dispersion near $E_{\rm F}$ and the FS topology.  The experimentally determined FS consists of two electron pockets centered at the M point.  The $\beta$ hole pocket seen in the superconducting BaFe$_{1.85}$Co$_{0.15}$As$_2$ sample is absent or very small, resulting in the suppression of the $Q_{\rm AF}$ = ($\pi$, 0) interband scattering through the electron pocket at the M point, which is likely responsible for the disappearance of $T_{\rm c}$ at this heavily electron doping level.

We thank X. Dai, Z. Fang, and Z. Wang for providing their band-calculation results and valuable discussions.  We also thank T. Kawahara for his assistance in the ARPES experiment.  This work was supported by grants from JSPS, JST-CREST, MEXT of Japan, the Chinese Academy of Sciences, NSF, Ministry of Science and Technology of China, and NSF of US.
 

\end{document}